\pdfoutput=1
\documentclass{JHEP3}
\usepackage{amsmath}
\usepackage{graphics}
\usepackage{epsfig}
\usepackage{mathdots}

\def\({\left(}
\def\){\right)}

\newcommand{\be}{\begin{equation}}
\newcommand{\ee}{\end{equation}}
\newcommand{\bal}{\begin{aligned}}
\newcommand{\eal}{\end{aligned}}

\newcommand{\labell}[1]{\label{#1}}
\title{Cyclicity of All Anti-NMHV and N$^2$MHV Tree Amplitudes in $\mathcal{N}\!=\!4$ SYM}

\author{Junjie Rao$^{a}$\footnote{Email: jrao@aei.mpg.de}\\
{$^a$Max Planck Institute for Gravitational Physics (Albert Einstein Institute), 14476 Potsdam, Germany}}

\abstract{This article proves the cyclicity of anti-NMHV and N$^2$MHV tree amplitudes in planar $\mathcal{N}\!=\!4$ SYM
up to any number of external particles as an interesting application of positive Grassmannian geometry.
In this proof the two-fold simplex-like structures of tree amplitudes introduced in 1609.08627 play a key role,
as the cyclicity of amplitudes will induce similar simplex-like structures for the boundary generators of
homological identities. For this purpose, we only need a part of all distinct boundary generators, and the relevant
identities only involve BCFW-like cells. The manifest cyclic invariance in this geometric representation
reflects one of the invariant characteristics of amplitudes, though they are obtained by the scheme-dependent BCFW
recursion relation.}

\keywords{Amplitudes, Positive Grassmannian}

\begin{document}
\maketitle

%%%%%%%%%%%%%%%%%%%%%%%%%%%%%%%%%%%%%%%%%%%%%%%%%%%%%%%%%%%%%%%%%%%%%%%%%%%%%%%%
%%%%%%%%%%%%%%%%%%%%%%%%%%%%%%%%%%%%%%%%%%%%%%%%%%%%%%%%%%%%%%%%%%%%%%%%%%%%%%%%
%%%%%%%%%%%%%%%%%%%%%%%%%%%%%%%%%%%%%%%%%%%%%%%%%%%%%%%%%%%%%%%%%%%%%%%%%%%%%%%%
\section{Introduction}

$\mathcal{N}\!=\!4$ super Yang-Mills theory has been the most understood quantum field theory so far.
In recent years, tremendous progress on the scattering amplitudes of planar $\mathcal{N}\!=\!4$ SYM was made through
its connection to positive Grassmannian, momentum twistors and on-shell diagrams
\cite{ArkaniHamed:2012nw,ArkaniHamed:2009vw}. In this new context, the well known BCFW recursion
relation \cite{Britto:2005fq} using massless spinors has an elegant generalization \cite{ArkaniHamed:2010kv}
in momentum twistor space. This efficient machinery is powerful for generating tree amplitudes and loop integrands,
and it can manifest dual superconformal invariance of planar $\mathcal{N}\!=\!4$ SYM.
There is a \textsc{Mathematica} package ``\verb"positroids"'' to implement these results, with investigations on
various mathematical aspects of positive Grassmannian \cite{Bourjaily:2012gy}.
More relevant background on amplitudes can be found in \cite{Elvang:2013cua,Henn:2014}.

In particular, tree amplitudes in planar $\mathcal{N}\!=\!4$ SYM have an impressive simplicity in the language
of positive Grassmannian in momentum twistor space, namely the so-called two-fold simplex-like structures
\cite{Rao:2016out}, a concise review can be referred in \cite{Rao:2017sbs}. In terms of Grassmannian geometry
representatives specifying linear dependencies of different ranks and empty slots for null columns, information of
amplitudes can be compactly captured by finite numbers of fully-spanning cells and their growing parameters.
Given a fixed $k$, as $(k\!+\!2)$ is the number of negative helicities, there is no new full cell beyond $n\!=\!4k\!+\!1$,
then after we identify all full cells with their growing parameters at this critical $n$, N$^k$MHV
amplitudes are known once for all up to any number of external particles.
This is an extension following the logic similar to \cite{Drummond:2008cr,ArkaniHamed:2009dg}.

With the aid of this purely geometric description, homological identities can be understood in a much more intuitive
way, and most of them turn out to be the secret incarnation of the simple NMHV identity.
A part of these identities are crucial for interconnecting different BCFW cells, and hence different BCFW recursion
schemes \cite{ArkaniHamed:2012nw}. Explicitly in this work, we would like to manifest the cyclicity of
amplitudes of two specific classes: the anti-NMHV and the N$^2$MHV families, by applying the simplex-like structures
of both the amplitudes and boundary generators of identities. From \cite{Rao:2016out} we have fully understood the structures
of anti-NMHV, NMHV, N$^2$MHV and N$^3$MHV families, while only the cyclicity of NMHV family and $n\!=\!7,8$ anti-NMHV
amplitudes has been shown. It is then desirable to see more nontrivial examples and attempt to extract the general pattern
from them. And the manifest cyclic invariance reflects one of the invariant characteristics of amplitudes, which is obscured
by the scheme-dependent BCFW recursion relation. This cyclicity is not manifest in the amplituhedron setting
\cite{Arkani-Hamed:2013jha,Arkani-Hamed:2017vfh,Kojima:2020tjf,Lukowski:2020bya} as well, as the triangulation process
usually chooses some fixed labels of particles for simplicity. Moreover, though the NMHV identity has an obvious geometric
interpretation as different triangulations give the same invariant sum of ``volumes'', for more general N$^k$MHV identities
with $k\!\geq\!2$ the corresponding geometric pictures are unclear yet. From the perspective of cyclicity, we may find more
intuition of these identities which interconnect BCFW cells of different Grassmannian geometric configurations.
We may even go further to find their counterparts in the context of amplituhedron, and in particular, explore their
relation to the sign-flip triangulation \cite{Arkani-Hamed:2017vfh,Kojima:2020tjf}.

As a helpful warmup exercise, we now reconsider the cyclicity of NMHV family in a more formal way before we derive
its generalization for $k\!\geq\!2$ in this work.

Recall the NMHV $n\!=\!6$ amplitude in terms of empty slots in the default recursion scheme is given by
\be
Y^1_6=[6]+[4]+[2],
\ee
where $Y^k_n$ is the Yangian invariant related to the $n$-particle amplitude with $(k\!+\!2)$ negative helicities,
via $A^k_n=A^{\textrm{MHV}}_nY^k_n$ (the MHV sector means $k\!=\!0$), and an empty slot $[i]$ of $k\!=\!1,n\!=\!6$
denotes the commonly used 5-bracket with entry $i$ removed \cite{ArkaniHamed:2012nw,Rao:2016out}.
Then the difference between $Y^1_6$ and its cyclicly shifted (by $+1$) counterpart is
\be
Y^1_6-Y^1_{6,+1}=-\,[1]+[2]-[3]+[4]-[5]+[6]\equiv I_{123456},
\ee
here the 6-term NMHV identity of labels $1,2,3,4,5,6$ is defined as $I_{123456}$.
According to the simplex-like structures of amplitudes, namely the quadratic mode with growing parameters $(6,4,2)$,
given by (in this triangle-shape sum each entry is multiplied by its corresponding vertical and horizontal factors)
\be
Y^1_n=
\(\begin{array}{c|cccccc}
[234\ldots n\!-\!4]~ & {} & {} & {} & {} & {} & 1 \\
\vdots~~ & {} & {} & {} & {} & \,\iddots & \vdots \\
{[234]}~ & {} & {} & {} & ~~~1 & ~\cdots & ~~~~[\ldots n\!-\!2] \\
{[23]}~~ & {} & {} & ~1 & ~~~[6] & ~\cdots & ~~~[6\ldots n\!-\!2] \\
{[\textbf{2}]}~~~ & {} & 1 & ~[5] & ~~[56] & ~\cdots & ~~[56\ldots n\!-\!2] \\
1~~~ & ~1 & [\textbf{4}] & [45] & ~[456] & ~\cdots & ~[456\ldots n\!-\!2] \\
\hline
{} & ~[\textbf{6}78\ldots n] & ~[78\ldots n] & ~[8\ldots n] & ~[\ldots n] & ~\cdots & 1
\end{array}\), \labell{eq-5}
\ee
we have the following relation for $n\!=\!7$ as an example:
\be
Y^1_7-Y^1_{7,+1}=
\(\begin{array}{ccc}
{} & {} & [23] \\
{} & [27] & [25] \\
{[67]} & [47] & [45]
\end{array}\)-
\(\begin{array}{ccc}
{} & {} & [34] \\
{} & [31] & [36] \\
{[71]} & [51] & [56]
\end{array}\)=
\(\begin{array}{cc}
{} & [3]\,I_{124567} \\
{[7]}\,I_{123456} & [5]\,I_{123467}
\end{array}\).
\ee
This already completes the proof of cyclicity for the general $Y^1_n$, since the growing parameters $(7,5,3)$ of
$I_{123456}$ have been identified, and as $n$ increases $I_{123456}$ also follows the simplex-like growing pattern.

We see that $(7,5,3)$ are closely related to $(6,4,2)$ of $Y^1_n$, which shows how
the cyclicity of amplitudes induces similar simplex-like structures for the relevant homological identities.
This intriguing feature will appear in a much more nontrivial form for N$^2$MHV amplitudes.

%%%%%%%%%%%%%%%%%%%%%%%%%%%%%%%%%%%%%%%%%%%%%%%%%%%%%%%%%%%%%%%%%%%%%%%%%%%%%%%%
%%%%%%%%%%%%%%%%%%%%%%%%%%%%%%%%%%%%%%%%%%%%%%%%%%%%%%%%%%%%%%%%%%%%%%%%%%%%%%%%
%%%%%%%%%%%%%%%%%%%%%%%%%%%%%%%%%%%%%%%%%%%%%%%%%%%%%%%%%%%%%%%%%%%%%%%%%%%%%%%%
\section{Cyclicity of Anti-NMHV Amplitudes}

Before moving to the N$^2$MHV family, let's first consider the cyclicity of all anti-NMHV amplitudes, since this is
in fact the nontrivial starting point for all N$^k$MHV cases. More explicitly, recall that for a given $k$ non-vanishing
amplitudes start with the anti-MHV sector $n\!=\!k\!+\!4$, which contains just one top cell, then the first interesting
case is the anti-NMHV sector $n\!=\!k\!+\!5$. It can be rearranged in the similar form of a triangle-shape sum
as its parity conjugate, namely the NMHV sector.

The anti-NMHV triangle-like pattern can be clearly observed in the series of examples below:
\be
Y^1_6=
\(\begin{array}{cc}
[2]\, &
\left\{\begin{array}{c}
\,~~[6] \\
\!{[4]}~~~
\end{array} \right.
\end{array}\!\),
\ee
\be
Y^2_7=
\(\begin{array}{ccc}
[2]~ &
(23)\left\{\begin{array}{c}
\!(67) \\
\!(45)~~~~~~
\end{array} \right. &
\!\!\!\!\!\!\left\{\begin{array}{c}
\,~~~[7] \\
\!(45)(71) \\
\!{[5]}~~~~
\end{array} \right.
\end{array}\!\!\),
\ee
\be
Y^3_8=
\(\begin{array}{cccc}
[2]~ &
(23)\left\{\begin{array}{c}
\!(678) \\
\!(456)~~~~~~
\end{array} \right. &
\!\!\!\!\!\!(234)\left\{\begin{array}{c}
~~~(78) \\
\!(456)(781)~~ \\
\!(56)~~~~~~~
\end{array} \right. &
\!\!\!\!\left\{\begin{array}{c}
~~~[8] \\
\!(456)(81)~~ \\
\,(56)(812) \\
\!\!{[6]}~~~~
\end{array} \right.
\end{array}\!\!\!\),
\ee
\be
Y^4_9=
\(\begin{array}{ccccc}
[2]~~ &
(23)\left\{\begin{array}{c}
\!(6789) \\
\!(4567)~~~~~~
\end{array} \right. &
\!\!\!\!(234)\left\{\begin{array}{c}
\,~~~~(789) \\
\!(4567)(7891)~~ \\
\!(567)~~~~~~~~~
\end{array} \right. &
\!(2345)\left\{\begin{array}{c}
\!(89) \\
\!(4567)(891)~~~~~~~ \\
\!(567)(8912)~~~~ \\
\!(67)~~~~~~~~~~~
\end{array} \right. &
\!\!\!\!\!\!\!\left\{\begin{array}{c}
\,~~~[9] \\
\!(4567)(91)~~~ \\
\!(567)(912) \\
\,~~(67)(9123) \\
\!{[7]}~~~~
\end{array} \right.
\end{array}\!\!\),
\ee
and its general form can be proved by induction. Here we remind that, for $k\!\geq\!2$, the empty slot $[i]$ again denotes
the Grassmannian cell with column $i$ removed, $(ij)$ denotes that columns $i,j$ are proportional and $(ijk),(ijkl)$
similarly denote linear dependencies of various ranks (in particular, $(i_1\ldots i_k)$ denotes a $k\!\times\!k$
vanishing minor) as defined in \cite{Rao:2016out}.

From \cite{Rao:2016out}, it is already known that
\be
Y^2_7-Y^2_{7,+1}=-\,\partial(23)-\partial(56)-\partial(71), \labell{eq-6}
\ee
\be
Y^3_8-Y^3_{8,+1}=\partial(23)+\partial(67)+\partial(81)+\partial(234)(567)+\partial(567)(812)+\partial(781)(234),
\ee
which manifest the cyclicity of $Y^2_7$ and $Y^3_8$. As boundary generators, the $(4k\!+\!1)$-dimensional cells above
lead to the relevant identities after the formal `$\partial$' operation. These results can be rearranged in a more
suggestive form as
\be
Y^2_7-Y^2_{7,+1}=-\,\partial
\(\begin{array}{cc}
(23)\, &
\left\{\begin{array}{c}
\!(71) \\
\!(56)~~~~~~
\end{array} \right.
\end{array}\!\!\!\!\!\!\!\), \labell{eq-1}
\ee
\be
Y^3_8-Y^3_{8,+1}=\partial
\(\begin{array}{cccc}
(23)~ &
(234)\left\{\begin{array}{c}
\!(781) \\
\!(567)~~~~~~
\end{array} \right. &
\!\!\!\!\!\!\!\left\{\begin{array}{c}
~~~(81) \\
\!(567)(812)~~ \\
\!\!(67)~~~~~~~
\end{array} \right.
\end{array}\!\!\!\!\!\!\), \labell{eq-2}
\ee
as well as a further $k\!=\!4$ extension of this pattern:
\be
Y^4_9-Y^4_{9,+1}=-\,\partial
\(\begin{array}{ccccc}
(23)~ &
(234)\left\{\begin{array}{c}
\!(7891) \\
\!(5678)~~~~~~
\end{array} \right. &
\!\!\!\!\!\!(2345)\left\{\begin{array}{c}
\,~~~~(891) \\
\!(5678)(8912)~~ \\
\!(678)~~~~~~~~~
\end{array} \right. &
\!\!\!\!\left\{\begin{array}{c}
\!(91) \\
\!(5678)(912)~~~~~~~ \\
\!(678)(9123)~~~~ \\
\!(78)~~~~~~~~~~~
\end{array} \right.
\end{array}\!\!\!\!\!\!\!\!\!\!\!\!\!\), \labell{eq-3}
\ee
where the sign factor $(-)^{k+1}$ for each of these relations follows the convention of
\cite{ArkaniHamed:2012nw,Bourjaily:2012gy}. And the types of homological identities used in
\eqref{eq-1} and \eqref{eq-2}, as already proved in \cite{Rao:2016out}, include
\be
\partial(23)=-\,[2]+[3]-(23)(45)+(23)(56)-(23)(67)+(23)(71) \labell{eq-4}
\ee
for $k\!=\!2$ (we have shifted boundary generator $(12)$ in \cite{Rao:2016out} to $(23)$,
and similar below), as well as
\be
\partial(23)=+\,[2]-[3]+(23)(456)-(23)(567)+(23)(678)-(23)(781),
\ee
\be
\bal
\!\!\!\!\!\!\!\!\!\!\!\!\!\!
\partial(234)(567)=&+(23)(567)-(34)(567)+(234)(56)-(234)(67)\\
&+(234)(567)(781)-(234)(567)(812)
\eal
\ee
for $k\!=\!3$, while those for $k\!=\!4$ used in \eqref{eq-3} are new, as given by
\be
\partial(23)=-\,[2]+[3]-(23)(4567)+(23)(5678)-(23)(6789)+(23)(7891),
\ee
\be
\bal
\!\!\!\!\!\!\!\!\!\!
\partial(234)(5678)=&-(23)(5678)+(34)(5678)-(234)(567)+(234)(678)\\
&-(234)(5678)(7891)+(234)(5678)(8912),
\eal
\ee
\be
\bal
\!\!\!\!\!\!\!\!\!\!
\partial(2345)(678)=&-(234)(678)+(345)(678)-(2345)(67)+(2345)(78)\\
&-(2345)(678)(8912)+(2345)(678)(9123),
\eal
\ee
\be
\bal
\partial(2345)(5678)(8912)=&-(234)(5678)(8912)+(345)(5678)(8912)-(2345)(567)(8912)\\
&+(2345)(678)(8912)-(2345)(5678)(891)+(2345)(5678)(912),
\eal
\ee
and they can be proved by using the similar matrix approach as done in \cite{Rao:2016out}. The examples of
anti-NMHV family again show how the cyclicity of amplitudes induces similar structures for the relevant identities,
and such 6-term identities for any $k$ can be easily guessed from the boundary generators then proved.

%%%%%%%%%%%%%%%%%%%%%%%%%%%%%%%%%%%%%%%%%%%%%%%%%%%%%%%%%%%%%%%%%%%%%%%%%%%%%%%%
%%%%%%%%%%%%%%%%%%%%%%%%%%%%%%%%%%%%%%%%%%%%%%%%%%%%%%%%%%%%%%%%%%%%%%%%%%%%%%%%
%%%%%%%%%%%%%%%%%%%%%%%%%%%%%%%%%%%%%%%%%%%%%%%%%%%%%%%%%%%%%%%%%%%%%%%%%%%%%%%%
\section{Cyclicity of N$^2$MHV Amplitudes}

Now we will start with the cyclicity of N$^2$MHV $n\!=\!7$ amplitude, namely \eqref{eq-6}, to explore its generalization
towards $n\!\geq\!8$. Recall the N$^2$MHV full cells along with their growing parameters are given by
\be
G_{7,0}=\left\{\begin{array}{c}
\!(45)(71) \\
\!{[5]}~~~~
\end{array} \right.~~~~~~~~~~~~~~~~~~~~~(5)~~~~~~
\ee
\be
G_{7,1}=(23)\left\{\begin{array}{c}
\!(67) \\
\!(45)~~~~~~
\end{array} \right.~~~~~~~~~~~~~~~(6,4)~~~
\ee
\be
G_{8,1}=\left\{\begin{array}{c}
\!(234)_2(678)_2~~~~~~~~~~~~~~~~~(7,4)~~ \\
\!(456)_2(781)_2~~~~~~~~~~~~~~~~~(7,5)~~ \\
\!(23)(456)_2(81)~~~~~~~~~~~~~~\,(6,4)~~
\end{array} \right.
\ee
\be
G_{9,2}=\left\{\begin{array}{c}
\!(2345)_2(6789)_2~~~~ \\
\!(23)(4567)_2(891)_2
\end{array} \right.~~~~~~~~~(8,6,4)
\ee
and for notational convenience, below we will suppress subscript `2' for consecutive vanishing $2\!\times\!2$ minors,
such as $(234)_2\!\equiv\!(234)\!=\!(23)(34)$, which is unambiguous as we restrict the discussion to the
N$^2$MHV sector from now on. To understand how the full cells capture the information of amplitudes up to any number
of external particles \cite{Rao:2016out}, let's give a brief review first.

According to the simplex-like structures of amplitudes similar to \eqref{eq-5}, for the general $Y^k_n$ we have
\be
Y^k_n=
\(\begin{array}{c|cccccc}
[234\ldots n\!-\!k\!-\!3]~ & {} & {} & {} & {} & {} & 1 \\
\vdots~ & {} & {} & {} & {} & ~\iddots~ & \vdots \\
{[234]}~ & {} & {} & {} & ~1 & ~\cdots~ & ~I_{n-3,4} \\
{[23]}~~ & {} & {} & 1 & ~I_{k+5,3} & ~\cdots~ & ~I_{n-2,3} \\
{[2]}~~~ & {} & 1 & I_{k+5,2} & ~I_{k+6,2} & ~\cdots~ & ~I_{n-1,2} \\
1~~~ & 1 & I_{k+5,1} & I_{k+6,1} & ~I_{k+7,1} & ~\cdots~ & I_{n,1}~~ \\
\hline
{} & ~[k\!+\!5~k\!+\!6~k\!+\!7\ldots n] & ~[k\!+\!6~k\!+\!7\ldots n] & ~[k\!+\!7\ldots n] & ~[\ldots n] & ~\cdots~ & 1
\end{array}\),
\ee
and only $I_{i,1}$ in the bottom row needs to be identified, while $I_{i,1+j}$ can be obtained by performing a partial
cyclic shift $i\!\to\!i\!+\!j$ except that label 1 is fixed, for all BCFW cells within $I_{i,1}$.

In the case of $k\!=\!2$, $I_{i,1}$ can be expressed in terms of $G_{i,m}$ above as
\be
\bal
I_{7,1}&=G_{7,0}+G_{7,1},\\
I_{8,1}&=G_{8,1}+(G_{7,0,2}+G_{7,1,2}),\\
I_{9,1}&=G_{9,2}+G_{8,1,2}+(G_{7,0,3}+G_{7,1,3}),\\
I_{10,1}&=G_{9,2,2}+G_{8,1,3}+(G_{7,0,4}+G_{7,1,4}),\\
I_{11,1}&=G_{9,2,3}+G_{8,1,4}+(G_{7,0,5}+G_{7,1,5}),
\eal
\ee
and so on, where $G_{i,m}$ is purely made of full cells with growing mode $m$,
namely these full cells have $(m\!+\!1)$ growing parameters, and the additional label $l$ in $G_{i,m,l}$ represents its
level during the simplex-like growth. Explicitly, up to level 3, $G_{7,0}\!\to\!G_{7,0,2}\!\to\!G_{7,0,3}$ of 0-mode
is given by
\be
\left\{\begin{array}{c}
\!(45)(71) \\
\!{[5]}~~~~
\end{array} \right.\!\!\to
[5]\left\{\begin{array}{c}
\!(46)(81) \\
\!{[6]}~~~~
\end{array} \right.\!\!\to
[56]\left\{\begin{array}{c}
\!(47)(91) \\
\!{[7]}~~~~
\end{array} \right.
\ee
and $G_{7,1}\!\to\!G_{7,1,2}\!\to\!G_{7,1,3}$ of 1-mode is given by
\be
\begin{array}{ccc}
(23)\left\{\begin{array}{c}
\!(67) \\
\!(45)~~~~~~
\end{array} \right. &
\!\!\!\!\!\!\!\!\to[6](23)\left\{\begin{array}{c}
\!(78) \\
\!(45)~~~~~~
\end{array} \right. &
\!\!\!\!\!\!\!\!\to[67](23)\left\{\begin{array}{c}
\!(89) \\
\!(45)~~~~~~
\end{array} \right. \\[+1.5em]
{} &
\!\!\!\!\!\!\!\!\to[4](23)\left\{\begin{array}{c}
\!(78) \\
\!(56)~~~~~~
\end{array} \right. &
\!\!\!\!\!\!\!\!\to[47](23)\left\{\begin{array}{c}
\!(89) \\
\!(56)~~~~~~
\end{array} \right. \\[+1.5em]
{} & {} &
\!\!\!\!\!\!\!\!\to[45](23)\left\{\begin{array}{c}
\!(89) \\
\!(67)~~~~~~
\end{array} \right.
\end{array}
\ee
and it is similar for $G_{8,1}$, finally $G_{9,2}\!\to\!G_{9,2,2}\!\to\!G_{9,2,3}$ of 2-mode is given by
\be
\begin{array}{ccc}
\left\{\begin{array}{c}
\!(2345)(6789)~~~~ \\
\!(23)(4567)(891)
\end{array} \right. &
\!\to[8]\left\{\begin{array}{c}
\!\!(2345)(679\,10)~~~~ \\
\!(23)(4567)(9\,10\,1)
\end{array} \right. &
\!\to[89]\left\{\begin{array}{c}
\!(2345)(67\,10\,\,11)~~~~ \\
\!(23)(4567)(10\,\,11\,1)
\end{array} \right. \\[+1.5em]
{} &
\!\to[6]\left\{\begin{array}{c}
\!\!(2345)(789\,10)~~~~ \\
\!(23)(4578)(9\,10\,1)
\end{array} \right. &
\!\to[69]\left\{\begin{array}{c}
\!(2345)(78\,10\,\,11)~~~~ \\
\!(23)(4578)(10\,\,11\,1)
\end{array} \right. \\[+1.5em]
{} &
\!\to[4]\left\{\begin{array}{c}
\!\!(2356)(789\,10)~~~~ \\
\!(23)(5678)(9\,10\,1)
\end{array} \right. &
\!\to[67]\left\{\begin{array}{c}
\!(2345)(89\,10\,\,11)~~~~ \\
\!(23)(4589)(10\,\,11\,1)
\end{array} \right. \\[+1.5em]
{} & {} &
\!\to[49]\left\{\begin{array}{c}
\!(2356)(78\,10\,\,11)~~~~ \\
\!(23)(5678)(10\,\,11\,1)
\end{array} \right. \\[+1.5em]
{} & {} &
\!\to[47]\left\{\begin{array}{c}
\!(2356)(89\,10\,\,11)~~~~ \\
\!(23)(5689)(10\,\,11\,1)
\end{array} \right. \\[+1.5em]
{} & {} &
\!\to[45]\left\{\begin{array}{c}
\!(2367)(89\,10\,\,11)~~~~ \\
\!(23)(6789)(10\,\,11\,1)
\end{array} \right. \\[+1.5em]
\end{array}
\ee
from which we can see that, knowing all full cells with their growing parameters at $n\!=\!10$ is sufficient for
generating $Y^2_n$ up to any $n$. Note the full cells (or fully-spanning cells) are named such
that none of their $i$ columns are removed when they first show up in $I_{i,1}$,
while $[5]$ in $G_{7,0}$ is an exception, as it is a descendent of the N$^2$MHV top cell at $n\!=\!6$ but put together
with $(45)(71)$ for convenience.

Given the summary above, the N$^2$MHV $n\!=\!8$ amplitude is
\be
Y^2_8=S_{8,2}+S_{8,1}+S_{8,0},
\ee
where we have separated the terms containing $2,1,0$ empty slots respectively by defining
\be
S_{8,2}=
\(\begin{array}{ccc}
{} & {} & [23] \\
{} & [28] & [26] \\
{[78]} & [58] & [56]
\end{array}\),~~~
S_{8,1}=
\(\begin{array}{cc}
{} & [2](56)(81)~~~~ \\[+0.4em]
{} &
[2](34)\left\{\begin{array}{c}
\!(78) \\
\!(56)
\end{array} \right. \\[+1.3em]
[8](45)(71)~~~~ & [5](46)(81)~~~~ \\[+0.4em]
[8](23)\left\{\begin{array}{c}
\!(67) \\
\!(45)
\end{array} \right. &
[6](23)\left\{\begin{array}{c}
\!(78) \\
\!(45)
\end{array} \right. \\[+1.5em]
{} &
[4](23)\left\{\begin{array}{c}
\!(78) \\
\!(56)
\end{array} \right.
\end{array}\!\!\),~~~
S_{8,0}=\left\{\begin{array}{c}
\!(234)(678)~~~~ \\
\!(456)(781)~~~~ \\
\!(23)(456)(81)
\end{array} \right.\!\!,
\ee
as indicated by the second subscript $i$ of $S_{n,i}$. Now the cyclicity of $Y^2_8$ is separated into three parts:
\be
Y^2_8-Y^2_{8,+1}=(S_{8,2}-S_{8,2,+1})+(S_{8,1}-S_{8,1,+1})+(S_{8,0}-S_{8,0,+1}),
\ee
where the third subscript `+1' similarly denotes the cyclic shift. Straightforwardly we find
\be
S_{8,2}-S_{8,2,+1}=
\left\{\begin{array}{c}
[8]\,(-\,\partial(23)-\partial(56)-\partial(71)) \\
{[6]}\,(-\,\partial(23)-\partial(57)-\partial(81)) \\
{[3]}\,(-\,\partial(24)-\partial(67)-\partial(81))
\end{array} \right.\Bigg|_{\,1} ,
\ee
and `$\,|_{\,1}$' denotes the truncation that only keeps terms containing one empty slot. For example,
\eqref{eq-4} can be separated as
\be
\partial(23)\,|_{\,1}=-\,[2]+[3],~~\partial(23)\,|_{\,0}=-\,(23)(45)+(23)(56)-(23)(67)+(23)(71).
\ee
Knowing growing parameters $(8,6,3)$ of $n\!=\!7$ boundary generators, or identities
$(-\partial(23)\!-\!\partial(56)\!-\!\partial(71))$, we can denote this result as
($B_n$ stands for boundary generators first induced by the cyclicity of $Y^2_n$)
\be
B_7=-\,(23)-(56)-(71)~~~~~~(8,6,3)
\ee
so that
\be
S_{8,2}-S_{8,2,+1}=(\partial B_7\,|_{\,1})_1,~~
S_{8,1}-S_{8,1,+1}=(\partial B_7\,|_{\,0})_1+\partial B_8\,|_{\,1},~~
S_{8,0}-S_{8,0,+1}=\partial B_8\,|_{\,0},
\ee
where $(\partial B_7\,|_{\,1})_1$ denotes the counterpart of $\partial B_7\,|_{\,1}$ when $n$ increases from 7 to 8,
according to its simplex-like growing pattern. In this way, we can figure out $\partial B_8\,|_{\,1}$ (and hence $B_8$)
via the second relation above,
\\ \\ \\ \\
namely (in the 5th and 8th lines below we have added $+[4](56)(81)$ and $-[4](56)(81)$ respectively)
\be
\bal
\partial B_8\,|_{\,1}=\,&\,(S_{8,1}-S_{8,1,+1})-(\partial B_7\,|_{\,0})_1\\
=\,&-[7](81)(34)+[8](71)(34)-[1](78)(34)\\
&+[2](34)(56)-[3](24)(56)+[4](23)(56)\\
&+[2](34)(78)-[3](24)(78)+[4](23)(78)\\
&-[3](56)(81)+[4](56)(81)-[5](34)(81)+[6](34)(81)+[8](34)(56)-[1](34)(56)\\
&+[8](12)(56)-[1](82)(56)+[2](81)(56)\\
&-[5](67)(34)+[6](57)(34)-[7](56)(34)\\
&-[4](56)(81)+[5](46)(81)-[6](45)(81)\\
=\,&\,\partial\,(+\,(781)(34)-(234)(56)-(234)(78)+(34)(56)(81)+(567)(34)-(812)(56)+(456)(81))\,|_{\,1}.
\eal
\ee
After identifying $B_8$ (by trial and error), we can check the third relation above, as
\be
\bal
\partial B_8\,|_{\,0}=\,&+(781)(234)-(781)(345)+(781)(34)(56)\\
\,&-(234)(567)+(234)(56)(78)-(234)(56)(81)\\
\,&-(234)(56)(78)+(234)(678)-(234)(781)\\
\,&-(34)(567)(81)-(34)(56)(781)+(34)(56)(812)+(234)(56)(81)\\
\,&+(567)(81)(34)-(567)(12)(34)+(567)(234)\\
\,&-(812)(34)(56)+(812)(456)-(812)(567)\\
\,&+(456)(781)-(456)(812)+(456)(81)(23)\\
=\,&\,S_{8,0}-S_{8,0,+1}
\eal
\ee
nicely obeys the required consistency. The $n\!=\!8$ identities used above can be referred in appendix \ref{app1}
and they can be classified into five distinct types.

Next, for the cyclicity of $Y^2_9$ we similarly have
\be
S_{9,2}-S_{9,2,+1}=(\partial B_7\,|_{\,0})_2+(\partial B_8\,|_{\,1})_1,~~
S_{9,1}-S_{9,1,+1}=(\partial B_8\,|_{\,0})_1+\partial B_9\,|_{\,1},~~
S_{9,0}-S_{9,0,+1}=\partial B_9\,|_{\,0},
\ee
\\ \\ \\ \\ \\
where the simplex-like growing patterns give
\be
S_{9,1}=
\(\begin{array}{cc}
[2]\left\{\begin{array}{c}
\!(345)(789)~~~~ \\
\!(567)(891)~~~~ \\
\!(34)(567)(91)
\end{array} \right. &
\left\{\begin{array}{c}
\![7](234)(689)~~~~ \\
\!{[4]}(235)(789)~~~~
\end{array} \right. \\[+1.5em]
{} &
\left\{\begin{array}{c}
\![7](456)(891)~~~~ \\
\!{[5]}(467)(891)~~~~
\end{array} \right. \\[+0.8em]
[9]\left\{\begin{array}{c}
\!(234)(678)~~~~ \\
\!(456)(781)~~~~ \\
\!(23)(456)(81)
\end{array} \right. &
\left\{\begin{array}{c}
\![6](23)(457)(91) \\
\!{[4]}(23)(567)(91)
\end{array} \right.
\end{array}\!\!\),~~~
S_{9,0}=\left\{\begin{array}{c}
\!(2345)(6789)~~~~ \\
\!(23)(4567)(891)
\end{array} \right.\!.
\ee
And via the first relation above we can figure out $(\partial B_8\,|_{\,1})_1$,
or the growing parameters of $B_8$, as
\be
\bal
B_8=&+(781)(34)-(234)(56)-(234)(78)+(34)(56)(81)~~~~~~(9,7,5,3)\\
&+(567)(34)~~~~~~~~~~~~~~~~~~~~~~~~~~~~~~~~~~~~~~~~~~~~~~~~~~~~~~~~~\,(9,8,6,3)\\
&-(812)(56)~~~~~~~~~~~~~~~~~~~~~~~~~~~~~~~~~~~~~~~~~~~~~~~~~~~~~~~~~\,(9,6,3)\\
&+(456)(81)~~~~~~~~~~~~~~~~~~~~~~~~~~~~~~~~~~~~~~~~~~~~~~~~~~~~~~~~~\,(9,6,4)
\eal
\ee
so that via the second relation we can similarly figure out $\partial B_9\,|_{\,1}$ (and hence $B_9$),
and the third one again serves as a consistency check. Explicitly, we find
\be
\bal
B_9=&-(2345)(789)-(7891)(345)\\
&+(8912)(567)-(5678)(234)-(5678)(91)(34)+(5678)(12)(34)\\
&-(912)(34)(567)+(4567)(891).
\eal
\ee
Following exactly the same logic, for the cyclicity of $Y^2_{10}$ we have
\be
\!\!\!
S_{10,2}-S_{10,2,+1}\!=\!(\partial B_8\,|_{\,0})_2+(\partial B_9\,|_{\,1})_1,~~
S_{10,1}-S_{10,1,+1}\!=\!(\partial B_9\,|_{\,0})_1+\partial B_{10}\,|_{\,1},~~
S_{10,0}-S_{10,0,+1}\!=\!\partial B_{10}\,|_{\,0},
\ee
where the simplex-like growing patterns give
\be
S_{10,1}=
\(\begin{array}{cc}
[2]\left\{\begin{array}{c}
\!(3456)(789\,10)~~~~\, \\
\!(34)(5678)(9\,10\,1)
\end{array} \right. &
[8]\left\{\begin{array}{c}
\!(2345)(679\,10)~~~~\, \\
\!(23)(4567)(9\,10\,1)
\end{array} \right. \\[+1.5em]
{} &
[6]\left\{\begin{array}{c}
\!(2345)(789\,10)~~~~\, \\
\!(23)(4578)(9\,10\,1)
\end{array} \right. \\[+1.5em]
\![10]\left\{\begin{array}{c}
\!(2345)(6789)~~~~ \\
\!(23)(4567)(891)
\end{array} \right. &
[4]\left\{\begin{array}{c}
\!(2356)(789\,10)~~~~\, \\
\!(23)(5678)(9\,10\,1)
\end{array} \right.
\end{array}\!\!\),
\ee
\\
note that $S_{10,0}\!=\!0$ since there is no new full cell at $n\!\geq\!10$ \cite{Rao:2016out},
and hence $\partial B_{10}\,|_{\,0}\!=\!0$. Explicitly, we find
\be
\bal
B_9=&-(2345)(789)-(7891)(345)~~~~~~~~~~~~~~~~~~~~~~~~~~~~~~~~~~~~~~~~~~~~~~~~~~~\,(10,8,5,3)\\
&+(8912)(567)-(5678)(234)-(5678)(91)(34)+(5678)(12)(34)~~~~~~(10,8,6,3)\\
&-(912)(34)(567)~~~~~~~~~~~~~~~~~~~~~~~~~~~~~~~~~~~~~~~~~~~~~~~~~~~~~~~~~~~~~~~~~~\,(10,7,5,3)\\
&+(4567)(891)~~~~~~~~~~~~~~~~~~~~~~~~~~~~~~~~~~~~~~~~~~~~~~~~~~~~~~~~~~~~~~~~~~~~~~\,(10,7,5)
\eal
\ee
as well as
\be
B_{10}=+\,(789\,10\,1)(3456)-(23456)(789\,10)+(5678)(9\,10\,1\,2)(34)+(45678)(9\,10\,1\,2)-(45678)(9\,10\,1)(23).
\ee
Finally, for the cyclicity of $Y^2_{11}$ we have
\be
\!\!\!\!\!\!\!\!
S_{11,2}-S_{11,2,+1}\!=\!(\partial B_9\,|_{\,0})_2+(\partial B_{10}\,|_{\,1})_1,~~
S_{11,1}-S_{11,1,+1}\!=\!(\partial B_{10}\,|_{\,0})_1+\partial B_{11}\,|_{\,1},~~
S_{11,0}-S_{11,0,+1}\!=\!\partial B_{11}\,|_{\,0},
\ee
and explicitly we find
\be
\bal
B_{10}=&+(789\,10\,1)(3456)-(23456)(789\,10)+(5678)(9\,10\,1\,2)(34)~~~~~~(11,9,7,5,3)\\
&+(45678)(9\,10\,1\,2)-(45678)(9\,10\,1)(23)~~~~~~~~~~~~~~~~~~~~~~~~~~~~~~(11,9,7,5)
\eal
\ee
which leads to $(\partial B_{10}\,|_{\,0})_1\!=\!0$,
and hence $\partial B_{11}\,|_{\,1}\!=\!0$. From $S_{11,0}\!=\!0$ we also have $\partial B_{11}\,|_{\,0}\!=\!0$,
therefore it is safe to conclude that $B_{11}\!=\!0$. We can summarize all these intriguing results as
\be
B_7=-\,(23)-(56)-(71)~~~~~~~~~~~~~~~~~~~~~~~~~~~~~~~~~~~~~~~~~~~~~~~~~~~~~~~~~~~~~~(8,6,3)~~~~
\ee
\be
\bal
B_8=&+(781)(34)-(234)(56)-(234)(78)+(34)(56)(81)~~~~~~~~~~~~~~~~~~~~~~(9,7,5,3)\\
&+(567)(34)~~~~~~~~~~~~~~~~~~~~~~~~~~~~~~~~~~~~~~~~~~~~~~~~~~~~~~~~~~~~~~~~~~~~~~~~~\,(9,8,6,3)~\,\\
&-(812)(56)~~~~~~~~~~~~~~~~~~~~~~~~~~~~~~~~~~~~~~~~~~~~~~~~~~~~~~~~~~~~~~~~~~~~~~~~~\,(9,6,3)~\,\\
&+(456)(81)~~~~~~~~~~~~~~~~~~~~~~~~~~~~~~~~~~~~~~~~~~~~~~~~~~~~~~~~~~~~~~~~~~~~~~~~~\,(9,6,4)~\,
\eal
\ee
\be
\bal
B_9=&-(2345)(789)-(7891)(345)~~~~~~~~~~~~~~~~~~~~~~~~~~~~~~~~~~~~~~~~~~~~~~~~~~~\,(10,8,5,3)\\
&+(8912)(567)-(5678)(234)-(5678)(91)(34)+(5678)(12)(34)~~~~~~(10,8,6,3)\\
&-(912)(34)(567)~~~~~~~~~~~~~~~~~~~~~~~~~~~~~~~~~~~~~~~~~~~~~~~~~~~~~~~~~~~~~~~~~~\,(10,7,5,3)\\
&+(4567)(891)~~~~~~~~~~~~~~~~~~~~~~~~~~~~~~~~~~~~~~~~~~~~~~~~~~~~~~~~~~~~~~~~~~~~~~\,(10,7,5)
\eal
\ee
\be
\bal
~~B_{10}=&+(789\,10\,1)(3456)-(23456)(789\,10)+(5678)(9\,10\,1\,2)(34)~~~~~~~~~~~~\,(11,9,7,5,3)\\
&+(45678)(9\,10\,1\,2)-(45678)(9\,10\,1)(23)~~~~~~~~~~~~~~~~~~~~~~~~~~~~~~~~~~~~\,(11,9,7,5)
\eal
\ee
which terminate at $n\!=\!10$ like the full cells. With $B_7,B_8,B_9,B_{10}$ and the growing parameters
of relevant boundary generators identified, the cyclicity of $Y^2_n$ for any $n$ is proved.
These identities are classified into $1,5,6,4$ distinct types with respect to $n\!=\!7,8,9,10$
in appendix \ref{app1}.

A final remark is, not all N$^2$MHV homological identities are required for this proof. Especially, those involving
the quadratic cell at $n\!=\!8$, namely $(12)(34)(56)(78)$, or the composite-linear cell at $n\!=\!9$, namely
$(123)(456)(789)$, are irrelevant.
These two non-BCFW-like cells will lead to extra non-unity factors along with the 5-brackets \cite{Rao:2016out},
which cannot be generated by recursion. Therefore it is desirable to find that they do not appear at all in the proof of
cyclicity for N$^2$MHV amplitudes, not even appear as canceling pairs in the intermediate steps.

Since the cyclicity of tree amplitudes can be divided into many sub-equalities in terms of homological identities,
the latter in fact have some kind of invariant meaning if we reshuffle an identity as ``terms with plus signs $=$
terms with minus signs''. It is definitely an interesting and geometrically profound direction to explore this more
refined invariance in the context of amplituhedron and sign flips \cite{Arkani-Hamed:2017vfh,Kojima:2020tjf}.
Especially, the $k\!\geq\!2$ identities should give nontrivial insights on the connection between positive
Grassmannian and the sign-flip triangulation. Of course, the cyclicity at higher $k$ will demand us to work out
identities up to $n\!=\!4k\!+\!2$ at least, for example $n\!=\!14$ for $k\!=\!3$, which is a straightforward but
very lengthy task, since a more transparent pattern for all $k$'s still awaits to be found.

\appendix
%%%%%%%%%%%%%%%%%%%%%%%%%%%%%%%%%%%%%%%%%%%%%%%%%%%%%%%%%%%%%%%%%%%%%%%%%%%%%%%%
%%%%%%%%%%%%%%%%%%%%%%%%%%%%%%%%%%%%%%%%%%%%%%%%%%%%%%%%%%%%%%%%%%%%%%%%%%%%%%%%
%%%%%%%%%%%%%%%%%%%%%%%%%%%%%%%%%%%%%%%%%%%%%%%%%%%%%%%%%%%%%%%%%%%%%%%%%%%%%%%%
\section{Relevant N$^2$MHV Homological Identities}
\label{app1}

Below we list all distinct N$^2$MHV homological identities that are relevant in this work.
Note that we have discarded boundary cells that fail to have kinematical supports in terms of momentum twistors,
but still we abuse the term ``homological'' here
while the actual kinematics also matters \cite{Rao:2016out,Rao:2017sbs}.\\ \\
$n\!=\!7$
\be
\partial(12)=-\,[1]+[2]-(12)(34)+(12)(45)-(12)(56)+(12)(67).
\ee
$n\!=\!8$
\be
\partial(123)(45)=-\,[1](23)(45)+[2](13)(45)-[3](12)(45)+(123)(456)-(123)(45)(67)+(123)(45)(78).
\ee
\be
\partial(123)(56)=-\,[1](23)(56)+[2](13)(56)-[3](12)(56)+(123)(456)-(123)(567)+(123)(56)(78).~~\,
\ee
\be
\partial(123)(67)=-\,[1](23)(67)+[2](13)(67)-[3](12)(67)+(123)(45)(67)-(123)(567)+(123)(678).~~~
\ee
\be
\partial(123)(78)=-\,[1](23)(78)+[2](13)(78)-[3](12)(78)+(123)(45)(78)-(123)(56)(78)+(123)(678).
\ee
\be
\bal
\!\!\!\!\!\!\!
\partial(12)(34)(67)=&-[1](34)(67)+[2](34)(67)-[3](12)(67)+[4](12)(67)+[6](12)(34)-[7](12)(34)\\
&-(12)(345)(67)-(12)(34)(567)+(12)(34)(678)+(812)(34)(67).
\eal
\ee
$n\!=\!9$
\be
\bal
\partial(1234)(567)=&-[1](234)(567)+[2](134)(567)-[3](124)(567)+[4](123)(567)\\
&-(1234)(5678)+(1234)(567)(89).
\eal
\ee
\be
\bal
\partial(1234)(678)=&-[1](234)(678)+[2](134)(678)-[3](124)(678)+[4](123)(678)\\
&-(1234)(5678)+(1234)(6789).
\eal
\ee
\be
\bal
\partial(1234)(789)=&-[1](234)(789)+[2](134)(789)-[3](124)(789)+[4](123)(789)\\
&-(1234)(56)(789)+(1234)(6789).
\eal
\ee
\be
\bal
\partial(1234)(56)(89)=&-[1](234)(56)(89)+[2](134)(56)(89)-[3](124)(56)(89)+[4](123)(56)(89)\\
&-(1234)(567)(89)+(1234)(56)(789).
\eal
\ee
\be
\bal
\partial(1234)(67)(89)=&-[1](234)(67)(89)+[2](134)(67)(89)-[3](124)(67)(89)+[4](123)(67)(89)\\
&-(1234)(567)(89)+(1234)(6789).
\eal
\ee
\be
\bal
\partial(123)(45)(678)=&-[1](23)(45)(678)+[2](13)(45)(678)-[3](12)(45)(678)\\
&+[4](123)(678)-[5](123)(678)+[6](123)(45)(78)\\
&-[7](123)(45)(68)+[8](123)(45)(67)\\
&+(123)(45)(6789)-(9123)(45)(678).
\eal
\ee
$n\!=\!10$
\be
\bal
\!\!\!\!
\partial(12345)(6789)=&-[1](2345)(6789)+[2](1345)(6789)-[3](1245)(6789)\\
&+[4](1235)(6789)-[5](1234)(6789)+(12345)(6789\,10).
\eal
\ee
\be
\bal
\partial(12345)(789\,10)=&-[1](2345)(789\,10)+[2](1345)(789\,10)-[3](1245)(678\,10)\\
&+[4](1235)(789\,10)-[5](1234)(789\,10)+(12345)(6789\,10).
\eal
\ee
\be
\bal
\partial(12345)(678)(9\,10)=&-[1](2345)(678)(9\,10)+[2](1345)(678)(9\,10)-[3](1245)(678)(9\,10)\\
&+[4](1235)(678)(9\,10)-[5](1234)(678)(9\,10)+(12345)(6789\,10).
\eal
\ee
\be
\bal
\partial(1234)(5678)(9\,10)=&-[1](234)(5678)(9\,10)+[2](134)(5678)(9\,10)-[3](124)(5678)(9\,10)\\
&+[4](123)(5678)(9\,10)-[5](1234)(678)(9\,10)+[6](1234)(578)(9\,10)\\
&-[7](1234)(568)(9\,10)+[8](1234)(567)(9\,10)\\
&-[9](1234)(5678)+[10](1234)(5678).
\eal
\ee
%

%%%%%%%%%%%%%%%%%%%%%%%%%%%%%%%%%%%%%%%%%%%%%%%%%%%%%%%%%%%%%%%%%%%%%%%%%%%%%%%%
%%%%%%%%%%%%%%%%%%%%%%%%%%%%%%%%%%%%%%%%%%%%%%%%%%%%%%%%%%%%%%%%%%%%%%%%%%%%%%%%
%%%%%%%%%%%%%%%%%%%%%%%%%%%%%%%%%%%%%%%%%%%%%%%%%%%%%%%%%%%%%%%%%%%%%%%%%%%%%%%%

\end{document}